\documentclass{webofc}

\usepackage[varg]{txfonts}   
\usepackage{hyperref}
\usepackage{url}
\hypersetup{colorlinks=true,citecolor=blue,urlcolor=blue,linkcolor=blue}
\begin{document}
\title{Future facilities: the CERN SPS}
%
%

\author{\firstname{Roberta} \lastname{Arnaldi}\inst{1}\fnsep\thanks{\email{arnaldi@to.infn.it}} }

\institute{INFN Sezione di Torino, Via P. Giuria 1, 10125 Torino, Italy}

\abstract{The Super Proton Synchrotron (SPS) at CERN has played a pioneering role in the study of heavy-ion collisions since 1986 and nowadays remains central to the exploration of the Quark Gluon Plasma. This document summarizes the present status and future prospects of the SPS physics program with particular focus on hard and electromagnetic probes, highlighting the results and goals of NA61/SHINE and the proposed NA60+ experiment.
}
\maketitle
\section{Introduction: the CERN SPS}
\label{intro}
The Super Proton Synchrotron (SPS) at CERN began operation in 1976 with the acceleration of its first proton beams. Originally designed as a high-energy proton synchrotron, the SPS has since then become a versatile and essential component of CERN’s accelerator complex. Over the decades, it has been adapted to accelerate not only protons, but also antiprotons, electrons,  positrons and heavy-ions. The SPS has played a central role as an injector, first for the Large Electron–Positron Collider (LEP) and, more recently, for the Large Hadron Collider (LHC). However, beyond its injector role, the SPS has always supported a broad physics program, including fixed-target experiments and test beams, making it a key facility for both fundamental research and experimental development in high-energy physics.

In this document, we will highlight to the role of SPS as an accelerator for heavy ions. Since 1986, in fact, different types of ions have been accelerated over a wide energy range, between 6 $<\sqrt{s_{NN}}<$ 17 GeV, reaching beam intensities as high as $10^7$ $s^{-1}$ and supporting numerous beam lines and experimental areas, providing beams to fixed-target experiments.  

The SPS has a long and successful history in advancing heavy-ion physics. Early experiments such as NA38, NA50~\cite{NA50:2004sgj}, NA60~\cite{NA60:2008dcb,NA60:2006ymb}, and CERES~\cite{CERES:2005uih} revealed key signatures of the Quark-Gluon Plasma (QGP), including the suppression of quarkonia, the modifications of vector meson spectral functions and the first evidence of temperatures, measured via thermal dimuons, higher than the critical temperature. 

While much has been learned at the top SPS energies, significant regions of the QCD phase diagram, especially at high baryochemical potential ($\mu_B$), remain to be explored.
As heavy-ion physics enters a new phase with more precise and differential probes, the SPS remains an essential facility for studying rare processes in the intermediate energy range between SIS100/FAIR and RHIC/BNL. This energy range is particularly suited for investigating the QCD phase transition, chiral symmetry restoration, and thermal radiation. 

In the following, the status and the future perspectives of the NA61/SHINE experiment and of the proposed NA60+, both at SPS,  will be discussed. 

\section{The NA61/SHINE experiment}

NA61/SHINE is a fixed-target experiment designed for a broad physics program, including hadron production, neutrino and cosmic ray reference measurements, and heavy-ion physics (see talk of A. Merzlaya at this conference and Ref.~\cite{Gazdzicki:995681,NA61:2014lfx}). Since 2009, it has conducted a systematic scan in energy and system size to search for the onset of deconfinement and signs of the QCD critical point.

\begin{figure}[h]
\centering
\includegraphics[width=5cm,clip]{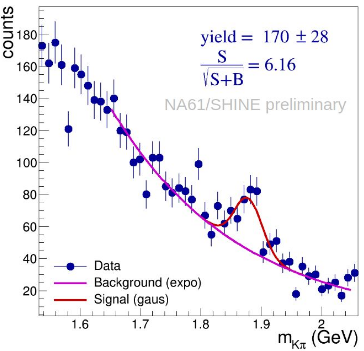}
\includegraphics[width=7.5cm,clip]{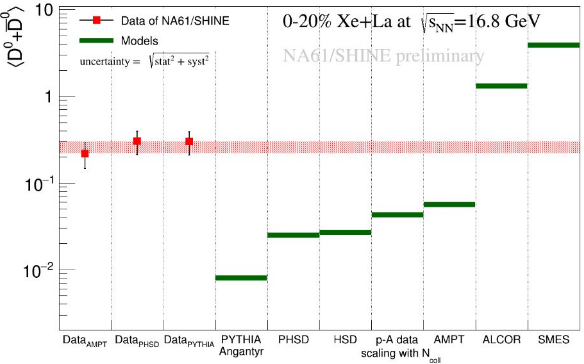}
\caption{Invariant mass spectrum, showing the D meson peak, reconstructed in the $K\pi$ decay channel, in central Xe-La collisions at $\sqrt{s_{NN}} = 16.8$ GeV (left panel); $<D^0+\bar{D}^0>$ yield compared to several theory models (right panel).}
\label{fig:NA61_charm}       
\end{figure}

One of the major recent results is the first direct observation of open charm production in Pb--Pb collisions at SPS energies. A clear $D^0$ signal has been measured, in its $K\pi$ decay channel with a statistical significance greater than $5\sigma$ in central Xe-La collisions at $\sqrt{s_{NN}}=~16.8$ GeV, as it can be observed in the left panel of Fig.~\ref{fig:NA61_charm}. The charm yield is compared to different model predictions, which span over up to two order of magnitude. In particular, the yields turn out to be significantly higher than expectation from proton-nucleus scaling, as shown in the right panel of Fig.~\ref{fig:NA61_charm}. Data precision, however, is not yet sufficient to discriminate between different theory models. 

The NA61/SHINE experimental setup has been significantly upgraded during the so-called LHC Long Shutdown 2 (LS2, taking place from 2019 to 2022). The new vertex telescope detector, equipped with ALPIDE sensors, together with the upgrade of the TPC readout electronics and DAQ, has guaranteed enhanced readout capabilities and improved vertexing, allowing NA61/SHINE to collect over 180 million Pb--Pb events at $\sqrt{s_{NN}} = 16.8$~GeV.

As future prospects, NA61/SHINE will continue its heavy-ion program with new beams (e.g., $^4$He, $^{10}$B, $^{16}$O), aiming to perform an energy scan (including beam energies of 13, 30 and 150 GeV) and to explore the deconfinement onset with light systems~\cite{Mackowiak-Pawlowska:2867952}. A proposed further upgrade includes replacing one of the time projection chambers (VTPC) with a fast silicon tracker to increase the event rate by a factor of 10, allowing new charm-correlation studies.

\section{The NA60+ Project}

NA60+ is a new experiment proposed at CERN SPS. NA60+, recently denominated DiCE, is dedicated to precision studies of the QCD phase diagram at high baryochemical potential.  Inspired by the success of the original NA60 experiment, NA60+ aims to conduct a detailed scan of Pb--Pb collisions in the range \( \sqrt{s_{NN}} \sim 6\)--17 GeV (i.e. $E_{lab} \sim 20 - 150$ GeV), corresponding to \( \mu_B \sim 220\)--550 MeV.


NA60+~\cite{NA60:2022sze} will perform precision studies of hard and electromagnetic probes, exploiting high-precision dimuon measurements and measuring the hadronic decays of charm hadrons and hypernuclei. Specifically, NA60+ will explore muon pairs from threshold up to $M_{\mu\mu}\simeq 4$~GeV/$c^{2}$, thereby enabling studies of the dilepton continuum, low-mass resonances, and quarkonium states. More details can be found in the presentation of A. Milov at this conference~\cite{Milov_HP}.

The main physics goals of NA60+ include:
\begin{itemize}
  \item investigation of chiral symmetry restoration via the study of the $\rho$-$a_1$ mixing.
  \item measurement of thermal radiation from the QGP through dimuon excess in the intermediate mass region.
  \item study of open charm hadrons and quarkonium production in a not yet explored high-$\mu_B$ environment.
  \item exploration of the nature of the phase transition.
\end{itemize}

\begin{figure}[h]
\centering
\includegraphics[width=9cm,clip]{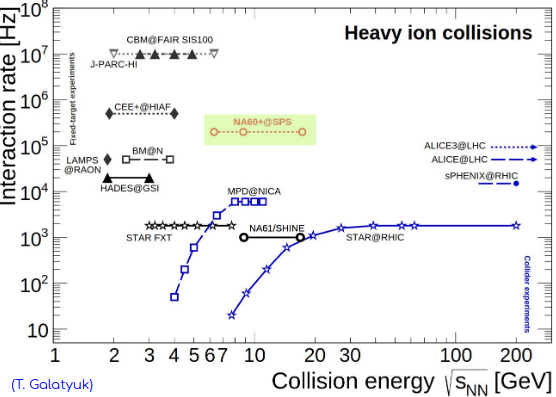}
\caption{Compilation of experiments complementary to NA60+ in terms of interaction rate and centre-of-mass energies (courtesy of T. Galatyuk).}
\label{fig:NA60+_complementarity}       
\end{figure}

NA60+ is uniquely positioned to access these observables with high precision due to the high luminosity available in fixed-target mode (up to $10^5$ interactions/s) and the capability to utilize Pb beams up to $10^6$/s across various energies. As illustrated in Fig.~\ref{fig:NA60+_complementarity}, NA60+ will be complementary to other current or planned experiments accessing different observables in a similar energy range, such as STAR BES~\cite{PhysRevC.107.L061901,201564}, NICA~\cite{Golovatyuk:2019rkb} or NA61~\cite{Gazdzicki:995681,NA61:2014lfx}, or similar observables in a lower energy range, such as CBM~\cite{Friman:2011zz}. 

\begin{figure}[h]
\centering
\includegraphics[width=12cm,clip]{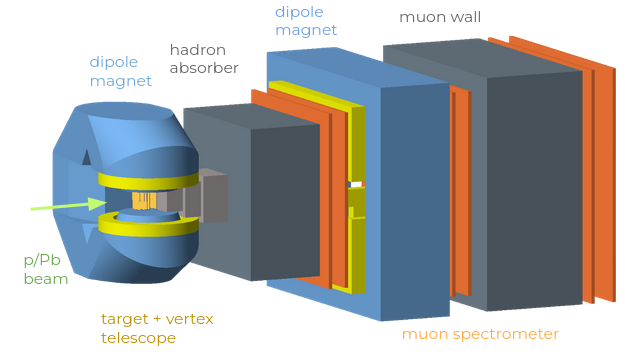}
\caption{Conceptual design of the NA60+ experimental setup~\cite{NA60:2022sze,NA60DiCE:2025qra}}
\label{fig:NA60+_setup}       
\end{figure}

The NA60+ experimental setup, inspired to the setup of the former NA60 experiment, is shown in Fig.~\ref{fig:NA60+_setup} and it comprises the following components:
\begin{itemize}
  \item a muon spectrometer located downstream of a hadron absorber, dedicated to the detection of dimuon pairs;
  \item a high-resolution silicon vertex telescope, located just downstream of the target system, providing precise vertex reconstruction and track measurements prior to the multiple scattering effects introduced by the absorber;
  \item two dipole magnets: one situated within the vertex telescope region and another placed between the stations of the muon spectrometer. These magnets allow accurate momentum determination for charged particle tracking both before and after the absorber.
\end{itemize}

\begin{figure}[h]
\centering
\includegraphics[width=8cm,clip]{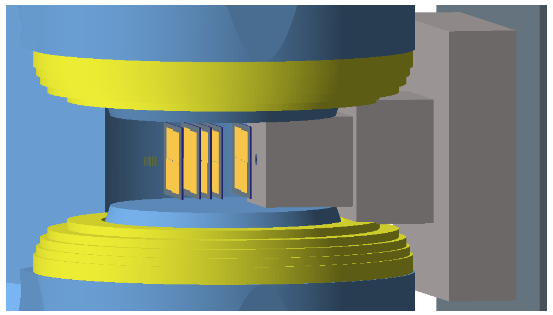}
\includegraphics[width=4.cm,clip]{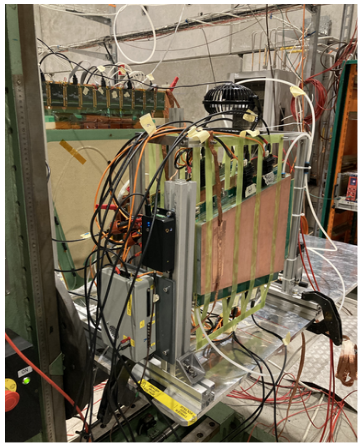}
\caption{NA60+ vertex telescope inside the MEP48 magnet (left panel); prototype of a muon spectrometer module, tested at CERN SPS in Fall 2023 (right panel)}
\label{fig:NA60+_setupzoom}       
\end{figure}

The vertex telescope of NA60+ comprises five stations, each of them from four large-area monolithic active pixel sensors (MAPS) measuring $13.6 \times 13.6$~cm$^2$. These sensors, developed in collaboration with the ALICE ITS3 project, are composed of seven MOSAIX segments, each formed by stitching together six basic units of 2.5~cm length. The sensors have a minimal thickness, contributing less than 0.1\% of a radiation length in the transverse direction, thereby reducing multiple scattering effects. A realistic sensor floorplan is currently available, and the final sensor prototype is foreseen in 2026. The total active area of the telescope is approximately 0.5~m$^2$. The entire vertex telescope is embedded within the 1.47~T dipole magnetic field provided by the MEP48 magnet, as illustrated in Fig.~\ref{fig:NA60+_setupzoom}.

The muon spectrometer of NA60+ will consist of four tracking stations designed to measure the positions of charged particle hits before and after the MNP33 dipole magnet, which has a 0.85 T·m integrated field. The recent availability of the MNP33 magnet, currently utilized by the NA62 experiment, allows for a significant cost reduction for NA60+ compared to the initially planned construction of a new toroidal magnet~\cite{NA60:2022sze}. Two additional stations, positioned downstream of a hadron absorber wall, will facilitate muon identification. The tracking stations will employ overlapping modules based on either Gas Electron Multiplier (GEM) or Multi-Wire Proportional Chamber (MWPC) technologies. The final design is currently under development. Initial prototypes of these chambers and their associated electronics have undergone beam tests at CERN, achieving spatial resolutions of approximately 0.1~mm in the $x$ direction and 0.5~mm in the $y$ direction, in line with the requirements of the NA60+ experiment.

This sophisticated detector system enables NA60+ to achieve very good mass resolution (better than 10 MeV/$c^2$ in the low mass region) and an excellent signal-to-background ratio, both of them essential for extracting rare electromagnetic probes in a high-multiplicity environment.

Significant attention in the NA60+ experiment is dedicated to the preparation of the Pb beam. The beam requirements are particularly stringent, necessitating a high-intensity beam of $10^7$ Pb ions per spill and a highly focused, sub-millimetric beam profile. This precision is essential to ensure the beam passes through a 6~mm diameter aperture in the vertex telescope planes without interacting with the sensitive detector areas. Beam optics studies are currently underway and results obtained from the first beam tests at SPS are promising.

\begin{figure}[h]
\centering
\includegraphics[width=5.5cm,clip]{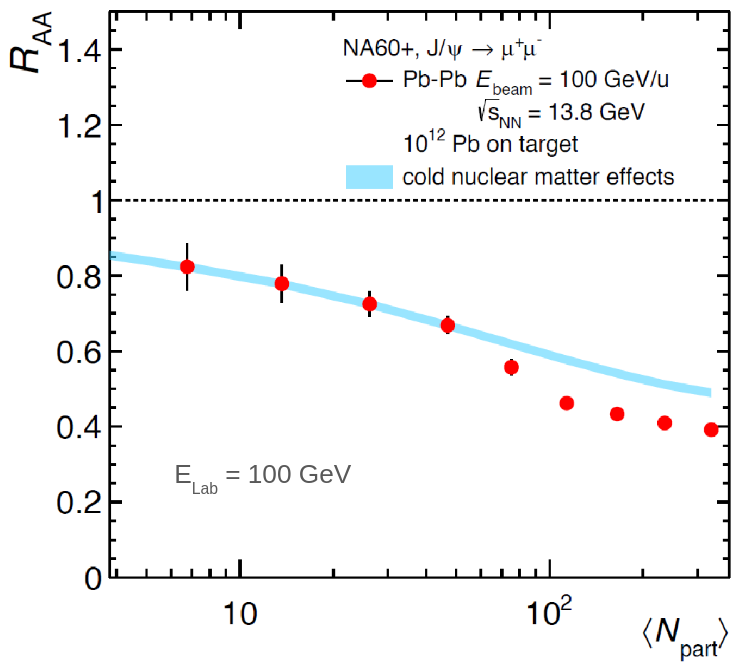}
\includegraphics[width=7cm,clip]{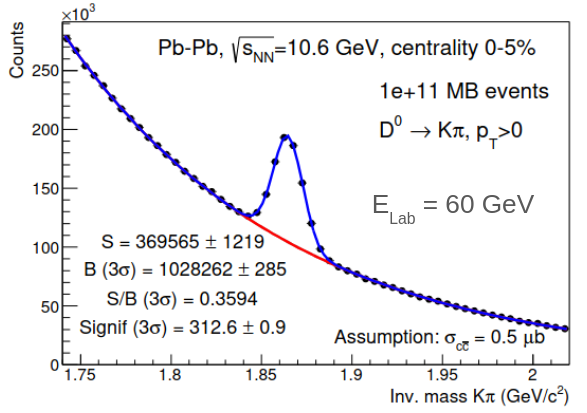}
\caption{$J/\psi$ nuclear modification factor expected at the beam energy of 100 GeV (left panel); $D^0$ invariant mass spectrum, in PbPb collisions at the energy of 60 GeV, obtained after collecting $10^{11}$ minimum bias events, corresponding to one month of data taking (right panel)}
\label{fig:NA60+_charm}       
\end{figure}

NA60+ is designed to deliver high-precision measurements across several key observables. The expected physics performance includes:
\begin{itemize}
  \item \textbf{Vector meson spectroscopy:} Thanks to its excellent mass resolution, NA60+ can accurately resolve the $\rho$, $\omega$, and $\phi$ peaks in the low mass region, making it possible to observe in-medium modifications of the $\rho$ spectral functions and assess chiral symmetry restoration. This will be achieved by studying the $\rho$-$a_1$ mixing, which should affect the dimuon continuum, resulting in a 20-30\% enhancement expected in the region $0.8 < M_{\mu\mu}< 1.5 GeV/c^2$
  \item \textbf{Thermal dimuons:} In the intermediate mass region, the precision dimuon measurement allows for the extraction of thermal radiation from the QGP. The temperature of the emitting source can be extracted from the inverse slope of the dimuon mass spectrum in the region $1.5 < M_{\mu\mu}< 2.5~GeV/c^2$. The measurement will have a $\sim1-3$\% uncertainty, allowing an accurate mapping of the $\sqrt{s_{NN}}$-dependence of the temperature, as visible in Fig.~\ref{fig:NA60+_T}~\cite{Galatyuk:2015pkq}. The achievable precision will also allow to have a strong sensitivity to a possible flattening of the caloric curve, due to a first-order phase transition.

\begin{figure}[h]
\centering
\includegraphics[width=9cm,clip]{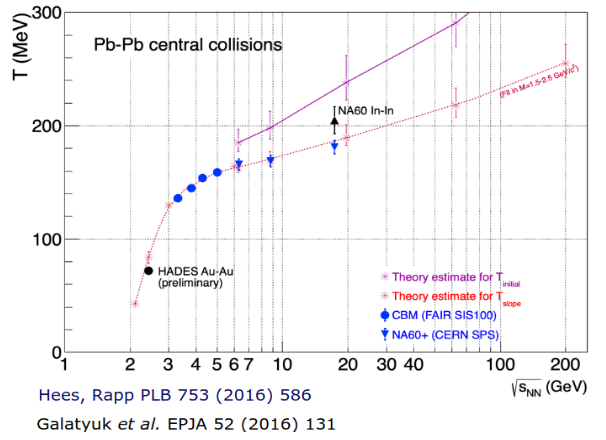}
\caption{$\sqrt{s_{NN}}$-dependence of the temperature. Existing results or predictions are also shown~\cite{Galatyuk:2015pkq}. }
\label{fig:NA60+_T}       
\end{figure}

  \item \textbf{Quarkonium measurements:} NA60+ will study quarkonium production for the first time at energies below the top SPS one. Furthermore, the foreseen proton-nucleus data takings will give access to cold nuclear matter effects, which might be relevant at low energy. The reachable statistics, of the order of $10^4-10^5$ $J/\psi$ will allow to study the nuclear modification factor $R_{AA}$ down to beam energies of the order of 50 GeV, as visible in the left panel of Fig.~\ref{fig:NA60+_charm}. Determining the onset of quarkonium suppression, it will be possible to correlate it with the temperature of the system measured via thermal dimuons, to identify the threshold temperature for the melting of the charmonium states.

  \item \textbf{Open Charm measurements:} The experiment is capable of measuring open charm, reconstructing their hadronic decays through topological cuts, as seen in the right panel of Fig.~\ref{fig:NA60+_charm} for the $D^0$ case. This will enable a clean measurement of charm production and its potential modification in a baryon-rich medium, accessing QGP transport properties, charm thermalization aspects, or hadronization mechanisms. In one month of data taking, the determination of the $D^0$ nuclear modification factor will be within reach, at all energies, even with $\sim$1\% statistical uncertainty, allowing $R_{AA}$ differential studies. Other open charm mesons or baryons (such has the $D_s$ or the $\Lambda_c$) will also be within reach, even if with a larger statistical precision.
\end{itemize}

All in all, the NA60+ represents as a powerful tool for mapping the phase diagram of strongly interacting matter and investigating the above-mentioned key phenomena.

The NA60+ experiment is currently preparing a proposal for submission to the CERN SPS Committee (SPSC) by mid-2025. Data taking is planned to begin after the LHC's Long Shutdown 3 (LS3), starting in 2029, and is expected to continue for approximately seven years. The experimental program includes proton-nucleus and lead-lead collisions at various energies, with each year dedicated to one energy point.

\section{Conclusions}

The SPS continues to be a vital facility for QCD studies at intermediate energies. NA61/SHINE has already demonstrated its unique capabilities for open charm and system-size scans. The proposed NA60+ experiment promises to deliver high-precision results on thermal radiation, chiral symmetry, and on open of hidden charm. Dusting off the glories of the first generation of SPS experiments, these efforts will provide deep insight into strongly interacting matter under extreme conditions, guiding future discoveries in heavy-ion physics, towards a bright SPS future!

\bibliography{HPtext} 

\begin{thebibliography}{15}

\bibitem{NA50:2004sgj}
B.~Alessandro et~al. (NA50), {A New measurement of J/psi suppression in Pb-Pb
  collisions at 158-GeV per nucleon}, Eur. Phys. J. C \textbf{39}, 335 (2005),
  \texttt{hep-ex/0412036}. \doiwoc{10.1140/epjc/s2004-02107-9}

\bibitem{NA60:2008dcb}
R.~Arnaldi et~al. (NA60), {Evidence for the production of thermal-like muon
  pairs with masses above 1-GeV/c**2 in 158-A-GeV Indium-Indium Collisions},
  Eur. Phys. J. C \textbf{59}, 607 (2009), \texttt{0810.3204}.
  \doiwoc{10.1140/epjc/s10052-008-0857-2}

\bibitem{NA60:2006ymb}
R.~Arnaldi et~al. (NA60), {First measurement of the rho spectral function in
  high-energy nuclear collisions}, Phys. Rev. Lett. \textbf{96}, 162302 (2006),
  \texttt{nucl-ex/0605007}. \doiwoc{10.1103/PhysRevLett.96.162302}

\bibitem{CERES:2005uih}
G.~Agakichiev et~al. (CERES), {e+ e- pair production in Pb - Au collisions at
  158-GeV per nucleon}, Eur. Phys. J. C \textbf{41}, 475 (2005),
  \texttt{nucl-ex/0506002}. \doiwoc{10.1140/epjc/s2005-02272-3}

\bibitem{Gazdzicki:995681}
M.~Gazdzicki, Z.~Fodor, G.~Vesztergombi (NA49-future), Tech. rep., CERN, Geneva
  (2006), revised version submitted on 2006-11-06 12:38:20,
  \urlstyle{tt}\url{https://cds.cern.ch/record/995681}

\bibitem{NA61:2014lfx}
N.~Abgrall et~al. (NA61), {NA61/SHINE facility at the CERN SPS: beams and
  detector system}, JINST \textbf{9}, P06005 (2014), \texttt{1401.4699}.
  \doiwoc{10.1088/1748-0221/9/06/P06005}

\bibitem{Mackowiak-Pawlowska:2867952}
M.~Mackowiak-Pawlowska (NA61/SHINE), Tech. rep., CERN, Geneva (2023),
  \urlstyle{tt}\url{https://cds.cern.ch/record/2867952}

\bibitem{NA60:2022sze}
C.~Ahdida et~al. (NA60+), {Letter of Intent: the NA60+ experiment} (2022),
  \texttt{2212.14452}.

\bibitem{Milov_HP}
M.~Alexander (NA60+), {The NA60+ experiment at SPS} (2025), \texttt{these
  proceedings}.

\bibitem{PhysRevC.107.L061901}
M.I. Abdulhamid et~al. (STAR), {Measurements of dielectron production in
  $\mathrm{Au}+\mathrm{Au}$ collisions at $\sqrt{{s}_{NN}}=27$, 39, and 62.4
  GeV from the STAR experiment}, Phys. Rev. C \textbf{107}, L061901 (2023).
  \doiwoc{10.1103/PhysRevC.107.L061901}

\bibitem{201564}
L.~Adamczyk et~al., {Energy dependence of acceptance-corrected dielectron
  excess mass spectrum at mid-rapidity in Au+Au collisions at sNN=19.6 and 200
  GeV}, Physics Letters B \textbf{750}, 64 (2015).
  \doiwoc{https://doi.org/10.1016/j.physletb.2015.08.044}

\bibitem{Golovatyuk:2019rkb}
V.~Golovatyuk, V.~Kekelidze, V.~Kolesnikov, O.~Rogachevsky, A.~Sorin,
  {Multi-Purpose Detector to study heavy-ion collisions at the NICA collider},
  Nucl. Phys. \textbf{A982}, 963 (2019).
  \doiwoc{10.1016/j.nuclphysa.2018.10.082}

\bibitem{Friman:2011zz}
B.~Friman, C.~Hohne, J.~Knoll, S.~Leupold, J.~Randrup, R.~Rapp, P.~Senger, {The
  CBM physics book: Compressed baryonic matter in laboratory experiments},
  Lect. Notes Phys. \textbf{814}, 1 (2011). \doiwoc{10.1007/978-3-642-13293-3}

\bibitem{NA60DiCE:2025qra}
C.~Ahdida et~al. (NA60+/DiCE), {European Strategy for Particle Physics 2026:
  the NA60+/DiCE experiment at the SPS} (2025), \texttt{2503.23872}.

\bibitem{Galatyuk:2015pkq}
T.~Galatyuk, P.M. Hohler, R.~Rapp, F.~Seck, J.~Stroth, {Thermal Dileptons from
  Coarse-Grained Transport as Fireball Probes at SIS Energies}, Eur. Phys. J. A
  \textbf{52}, 131 (2016), \texttt{1512.08688}.
  \doiwoc{10.1140/epja/i2016-16131-1}

\end{thebibliography}
\end{document}